\documentclass[aps,prb,amsmath,amssymb,twocolumn,floatfix,10pt,showpacs]{revtex4-1}
\usepackage{graphicx}
\usepackage{hyperref}
\usepackage{color}

\newcommand{\bs}{\;\;\;\;\;}

\newcommand{\ve}{\mathbf}
\newcommand{\D}{{\rm d}}
\newcommand{\tr}{{\rm Tr}}

\begin{document}

\title{Decoherence of Majorana qubits by noisy gates}
\author{Manuel J. Schmidt}
\affiliation{Institute for Theoretical Solid State Physics, RWTH Aachen, Sommerfeldstra\ss e 26, 52056 Aachen, Germany}
\author{Diego Rainis}
\author{Daniel Loss}
\affiliation{Department of Physics, University of Basel, Klingelbergstrasse 82, 4056 Basel, Switzerland}
\date{\today}


\begin{abstract}
We propose and study a realistic model for the decoherence of topological qubits, based on Majorana fermions in one-dimensional topological superconductors. The source of decoherence is the fluctuating charge on a capacitively coupled gate, modeled by non-interacting electrons. In this context, we clarify the role of quantum fluctuations and thermal fluctuations and find that quantum fluctuations do not lead to decoherence, while thermal fluctuations do. We explicitly calculate decay times due to thermal noise and give conditions for the gap size in the topological superconductor and the gate temperature. Based on this result, we provide simple rules for gate geometries and materials optimized for reducing the negative effect of thermal charge fluctuations on the gate.
\end{abstract}

\maketitle

\section{Introduction}

The central idea of topological quantum computing (TQC)\cite{tqc1,tqc2,tqc3,tqc_review,kitaev} is to encode quantum information in a degenerate ground state. In this context, the word ``topological" means that one must use a system in which the ground state degeneracy cannot be lifted by any local and sufficiently weak perturbation.\footnote{This statement is only true if the perturbation does not involve additional degrees of freedom that are not described by the same topological field theory (see Ref. \onlinecite{budich_2011}).} Such a system is said to be topologically nontrivial. If it is prepared in one of its many possible ground states, it will stay in exactly this state as long as the system is only weakly perturbed. In other words, the information encoded in this way is free of decoherence. The manipulations of the quantum information are performed via relatively large changes of material parameters that may be controlled externally, for instance via voltages applied to electric gates. The parameter variations needed in order to perform a nontrivial operation are so huge that they practically cannot happen accidentally. Moreover, the manipulation does not depend on the details of the parameter changes, but only on their topological properties. Thus, small imperfections in the parameter changes are not supposed to lead to imperfections in the operation on the quantum information. All this is true in an ideal world in which the system stays in its degenerate ground state and never gets excited.

One realization of such a topologically nontrivial system is the topological superconductor (TSC)\cite{TSC1,TSC2,TSC3,kitaev}. As in any superconductor, a finite energy $\geq\Delta$ must be paid in order to excite the system above its ground state. However, the ground state itself can be degenerate in a TSC; in addition to the fermionic excitations with energies $\geq\Delta$, there are $M$ ``excitations'' with zero energy, with $M$ the number of pairs of spatially separated Majorana bound states (MBS). Thus, the ground state is $2^M$-fold degenerate. For nontrivial rotations in this degenerate ground state space, the localized MBS must be moved (braided) around each other; such a braiding requires to change the local material parameters by large amounts, large enough that they cannot be due to uncontrollable fluctuations.

Recently, the concept of decoherence-free TQC, based on Majorana bound states in TSCs, has been challenged by the authors of Refs.~\onlinecite{chamon_decay_rates_2011,budich_2011}. In these works it was emphasized that topological protection does not protect against all kinds of perturbations that might be present in a realistic experiment. In particular, the work of Budich {\it et al.}~\cite{budich_2011} investigates tunnel couplings of the Majorana bound states to nearby fermionic baths that are not gapped. Since the topological protection of Majorana based qubits is based on the conservation of fermionic parity (moving one single electron in a superconductor costs energy $\geq \Delta$), but tunneling between ungapped fermionic baths and Majorana states allows single electrons to enter the TSC at zero energy, the quantum information in such a system is lost on exponentially short timescales. Thus, non-superconducting tunnel contacts close to the Majorana bound states must be avoided.

Goldstein and Chamon,\cite{chamon_decay_rates_2011} on the other hand, consider gapped fermionic baths which are coupled to the Majorana bound states by a fluctuating bosonic field, such as, for instance, phonons. In their work it becomes clear that a decay of the information (encoded by the local fermion parity) into a fermionic bath with an excitation gap $\Delta$ can occur only if the spectrum of the bosonic field is non-zero at frequencies $\omega\sim\Delta$. In other words, the energy that is needed to excite the TSC must be provided by the environment. One can also express this statement in terms of the adiabaticity requirement of TQC: if the external parameters of the Hamiltonian describing the TSC (e.g., electric potentials) are changed non-adiabatically, i.e., faster than $\hbar/\Delta$, then the TSC will not remain in its ground state, which is, however, an inevitable requirement for TQC.

Thus, it is clear that in the real world, where the constituents of an experimental setup are subject to all sorts of random fluctuations, the TSC will eventually get excited and lose its quantum information. Thus, an explicit study of the effect of different kinds of fluctuations on the TSC is inevitable for judging the viability of the concept of TQC. In this context, Ref.~\onlinecite{chamon_decay_rates_2011} raised the issue of decoherence caused by quantum fluctuations of bosonic fields. In some instances, this work is interpreted to the effect that a bosonic quantum bath at zero temperature may cause decoherence of a topological qubit. It is one of the goals of the present work to clarify the role of quantum fluctuations, or, more explicitly, zero temperature fluctuations.

Another much more obvious source of decoherence is the thermal fluctuation of a bosonic field. It is usually stated that the problems caused by those thermal fluctuations are exponentially small $\sim \exp(-\Delta/T)$. However, the exponential dependence alone is not enough for judging the feasibility of the concept of TQC - the pre-factor also matters. 
To be concrete, a decay time of a femtosecond times an exponentially large factor might still be extremely short for the experimentally accessible range of temperatures. Thus, it is of utmost importance to actually calculate the decay time of a topological qubit. Goldstein and Chamon made a first step in this direction,\cite{chamon_decay_rates_2011} but they considered a generic model. Thus, no predictions could be made about the actual decay time but only about its functional form. In order to make explicit predictions about the feasibility, one must {\it leave the level of generic models and focus on a particular experimental setup.} This is the path we will go in this paper.

The present work is based on an explicit experimental setup and a realistic model for one particular decoherence process, which we think is 
least avoidable in this setup (apart from quasiparticle poisoning, see Refs.~\onlinecite{rainis_qp_poisoning,FuKanePRB2009,AumentadoPRL2004,FergusonPRL2006,ShawPRB2008,MartinisPRL2009,CorcolesAPL11}). This explicitness finally allows us to plug in numbers and to obtain bounds for various parameters, e.g., the temperature or the superconducting gap. The experimental setup we choose has been proposed by Alicea and coworkers in Ref.~\onlinecite{alicea_1d_braiding_2011}, namely the movable Majorana bound states in networks of TSC wires. One of the basic building blocks of this proposal is an array of electric gates by which individual sections in the wire network are tuned into or out of their topologically nontrivial phases. The Majorana bound states live at the phase boundaries so that a MBS can be moved by means of changing gate voltages. In Ref.~\onlinecite{alicea_1d_braiding_2011} these gates were assumed to be classical objects giving rise to non-fluctuating local potentials that can be changed as slowly as needed. We extend this model by allowing charge fluctuations on the gates and study their effect on the TSC.

The paper is organized as follows. In Sec. II we discuss the setup and basic concepts. In Sec. III we derive an extended model for a TSC, including charge fluctuations on a nearby gate. The resulting decay time of a Majorana-based qubit is calculated in Sec. IV. In Sec. V we discuss our results and finally conclude in Sec VI.

\section{Preliminaries, setup and assumptions}
In order to give a self-contained and pedagogical description of the decoherence process and to further motivate our analysis, we start with some general considerations.

\subsection{Local fermion parity}
All our arguments are based on a mean-field description of the proximity-induced superconductivity. Within this description, the number of electrons is not fixed, since the Hamiltonian contains pairing terms. In particular, this means that the ground state has no well defined number of electrons. However, in usual superconductors without ground state degeneracy, the ground state $\left|\Omega\right>$ has a well defined fermion parity. Changing the fermion parity requires to occupy bogoliubons, i.e. $\left|\Omega\right>\rightarrow b_k^\dagger \left|\Omega\right>$, and this always requires a finite amount of energy in a usual superconductor. In a TSC with ground state degeneracy, however, there are bogoliubons $b_{0,m}$ with zero energy. Of course, these zero energy bogoliubons also change the fermion parity. Thus, one half of the degenerate ground states in a TSC has even parity, while the other half has odd parity.

Each zero energy bogoliubon consists of two MBS, the wave functions of which are spatially localized. This means, at least on
the mean-field level, that changing the global parity by changing the occupation of one zero energy bogoliubon $b_{0,m}$ does not affect the parts of the TSC that are far away from the MBS which form $b_{0,m}$ itself. {\it The parity change affects the TSC only locally.}

In general the mean-field description of a TSC contains $M+K$ bogoliubons. $M$ of them have zero energy and are annihilated by the fermionic operators $b_{0,m}$. Each $b_{0,m}$ is composed of two MBS. The remaining $K\gg M$ states are finite-energy Bogoliubov excitations $b_{k}$ with energies $\epsilon_k\geq\Delta$. For each $b_{0,m}$ we define a spatial region $R_m$ in which the wave function of $b_{0,m}$ is localized. We further assume that the regions are pairwise non-overlapping may be specified by the local parity in the regions $R_m$. As long as the TSC is in one if its ground states, and the positions of the regions $R_m$ do not change dramatically (no braiding), the local parity in $R_m$ cannot be changed.  This is because a change of parity would require one bogoliubon to travel away from $R_m$, and that in turn would involve extended states with energies $\geq \Delta$, which are not available to the system. As a consequence, the topological superconductor remains in exactly the ground state it has been prepared in.

\subsection{Decoherence processes}

There are several different processes related to charge fluctuations on a gate that may lead to decoherence. For instance, (a) a gate that is far away from any MBS may excite 
from the superconducting ground state a pair of high-energy bogoliubons, which travel around and accidentally hit a MBS. This is effectively similar to the phenomenon of quasiparticle poisoning	.\cite{rainis_qp_poisoning,FuKanePRB2009,AumentadoPRL2004,FergusonPRL2006,ShawPRB2008,MartinisPRL2009,CorcolesAPL11} Another possible process (b) is the excitation of one single high-energy bogoliubon in one region $R_m$ and the departure of this single bogoliubon from $R_m$. Due to parity conservation, this process (b) must also change the occupation of the zero-energy bogoliubon and thereby destroy the quantum information.

While processes such as (a) are expected to depend on global geometric properties of the setup (how likely is it that an average bulk excitation finds a MBS?), process (b) only depends on the local conditions close to the MBS in question. Therefore, we focus on this process in this work. More explicitly, we focus on the excitation subprocess in region $R_m$ governed by a perturbation
\begin{equation}
\delta Q (b_{k}^\dagger b_{0,m}+b_k^\dagger b_{0,m}^\dagger) + \rm H.c.,
\end{equation}
where $\delta Q$ is a bosonic field that describes the charge fluctuation on the gate.\cite{chamon_decay_rates_2011} Here we see that an excitation of a single high-energy bogoliubon $b_k^\dagger$ (describing an extended bulk state of the TSC) inevitably changes the state of the zero energy bogoliubon in $R_m$. We furthermore assume that, once it got excited, the finite-energy excitation immediately leaves the region $R_m$ and therewith changes the local parity in $R_m$.

\section{The model}
In this section we motivate the model on the basis of which we study the decoherence of quantum information via excitations of bulk bogoliubons with finite energies ($\geq \Delta$). It consists of three parts. One describes the topological superconductor $H_{\rm TSC}$, one describes the electrons on the nearby gate $H_{\rm G}$, and one describes the coupling between the TSC and the gate $H_c$.

The TSC is modeled by an effective p-wave superconductor in tight-binding representation\cite{kitaev,tqc1}
\begin{equation}
H_{\rm TSC} = \sum_n \left[-\frac{\Delta}{2} c_{n+1}c_n  -\frac{t_0}{2} c^\dagger_{n+1}c_n +  {\rm H.c.} -\mu_n c_n^\dagger c_n\right],\label{kitaev_hamiltonian}
\end{equation}
where $c_n$ annihilates an electron at site $n$, $\Delta$ is the superconducting gap, $t_0$ is the hopping amplitude, and $\mu_n$ is the electro-chemical potential at site $n$. We will assume later that $t_0$ is larger than any other energy scale in the TSC. In this limit, our results are independent of $t_0$. In the proposal by Alicea {\it et al.,}\cite{alicea_1d_braiding_2011} the local electric potential on the wire is tuned via the mean charge on a nearby gate. However, the charge on such a metallic gate may fluctuate and these fluctuations may induce quasiparticle excitations in the TSC. Therefore, we include the gate in the modeling by assuming non-interacting electrons in $D$ dimensions with an effective mass $m^*$
\begin{equation}
H_{\rm G} = \sum_{\ve k} \frac{\hbar^2 \ve k^2}{2m^*} c^\dagger_{\ve k}c_{\ve k}.
\end{equation}
The operators $c_{\ve k}$ annihilate an electron with momentum $\ve k$ in the gate. The capacitive coupling between the gate and the wire is then given by
\begin{equation}
\tilde H_c = \frac{e^2}{4\pi\epsilon\epsilon_0} \sum_n \int_G \D^D\ve r \frac{\psi^\dagger(\ve r)\psi(\ve r) c_n^\dagger c_n}{|\ve r - \ve x_n|},
\end{equation}
with $e$ the electron charge, $\epsilon$ the dielectric constant of the insulator between the wire and the gate, $\epsilon_0$ the vacuum permittivity, $\ve x_n$ the spatial position of the $n$th site in the TSC, $\ve r$ is integrated over the gate volume, and $\psi(\ve r) = L^{-D/2}\sum_\ve k e^{i\ve k\cdot\ve r}c_{\ve k}$ the field operators of the gate electrons with $L\rightarrow\infty$ being the linear size of the gate. In order to make the gate-wire coupling more tractable, we approximate $\tilde H_c$ by assuming $|\ve r-\ve x_n| \simeq d$, with $d$ the typical gate-wire distance, by restricting the $\ve r$ integral to a $D$-dimensional Gaussian profile of width $l\sim d$, and by restricting the $n$ summation by a profile function $F_n$. We have
\begin{multline}
\tilde H_c \simeq \underbrace{\frac{e^2}{4\pi\epsilon\epsilon_0 d} }_{\lambda} \underbrace{ \int \D^D\ve r e^{-\ve r^2/l^2}\psi^\dagger(\ve r)\psi(\ve r)}_{\equiv Q}\\ \times \sum_n F_n c_n^\dagger c_n \equiv H_c.
\end{multline}
In this work we will typically assume $F_n$ to be unity at the left end of the TSC (small $n$) where the wave function of the left Majorana state is non-zero, and zero on the right end of the wire (large $n$). It will turn out to be irrelevant how exactly $F_n$ approaches zero for large $n$. Furthermore, we may separate the mean charge $\left<Q\right>$ from the charge fluctuation $\delta Q = Q - \left<Q\right>$ and absorb $\left<Q\right>$ in the chemical potential in $H_{\rm TSC}$, i.e., we drop this term in $H_c$ and have
\begin{equation}
H_c = \lambda \delta Q \sum_n F_n c^\dagger_nc_n,
\end{equation}
with $\lambda = e^2/4\pi\epsilon\epsilon_0 d$.

In the full model
\begin{equation}
H = H_{\rm TSC} + H_{\rm G} + H_c
\end{equation}
$H_c$ excites the TSC quasiparticles via charge fluctuations $\delta Q$, the dynamics of which are governed by $H_{\rm G}$. In order to quantify this quasiparticle excitation, it is useful to transform from the electron operators $c_n$ on the TSC to the corresponding bogoliubon $b_n$. For simplicity we restrict the discussion to the case of $\mu=0$ and an odd number of TSC sites $N$ in the main text and show in Appendix \ref{nonzero_mu_gw} that these assumptions are not crucial for our results. After the Bogoliubov transformation one may write\footnote{We drop here one branch of the bogoliubon solutions which is disconnected for $\mu=0$. For details see Appendix \ref{nonzero_mu_gw}.}
\begin{equation}
H_{\rm TSC} = \sum_{k\neq0}\epsilon_k b^\dagger_kb_k \label{tsc_hamiltonian}
\end{equation}
with the quasiparticle excitation energies
\begin{equation}
\epsilon_k = \sqrt{ t_0^2\cos^2(k/2) + \Delta^2\sin^2(k/2)}.\label{bog_energy}
\end{equation}
Note, however, that there is an additional zero-energy quasiparticle $b_0$ that does not appear in $H_{\rm TSC}$. $b_0 = \gamma_L+i\gamma_R$ is composed of the left/right end MBS $\gamma_{L/R}$. Obviously, the ground state of $H_{\rm TSC}$ is two-fold degenerate, with $b_0^\dagger b_0 = 0,1$. Furthermore, $k=2\pi m/(N+1),\;\; m=1,2,... , (N-1)/2$.

The gate-wire coupling $H_{c}$, expressed in terms of the Bogoliubov quasiparticles, is of the form
\begin{equation}
H_c = \lambda \delta \hat Q \sum_{k, k'} \left[ b^\dagger_k X_{kk'} b_{k'} + \frac12 (b_k Y_{kk'} b_{k'} +{\rm H.c.}) \right] , \label{gw_coupling_bogol}
\end{equation}
where the sum over $k,k'$ includes the finite-energy as well as the zero-energy bogoliubons. The terms containing $b_0$ and $b_k$ with $k>0$ describe the excitation of a localized zero-energy bogoliubon to a delocalized high-energy bulk bogoliubon, while terms containing only bulk bogoliubons $b_k$ for $k>0$ describe the scattering between bulk bogoliubons or the excitation of a pair of bulk bogoliubons. Since we are only interested in the processes that involve the Majorana state, we keep only those terms involving $b_0$ operators. The corresponding coupling amplitudes read
\begin{multline}
X_{0k} = X_{k0} = Y_{0k} = -Y_{k0} \\
= - \frac{\sqrt{1-\delta^2}}{\sqrt{N+1}} \frac{\sin(k + \alpha_k) - \delta \sin(\alpha_k)}{1+\delta^2-2\delta\cos(k)}, \label{maj_bulk_coupling}
\end{multline}
with $\delta = (\Delta-t_0)/(\Delta+t_0)$ and
\begin{equation}
\alpha_k = -\arccos\left(\frac{ t_0\cos^2(k/2)+ \Delta\sin^2(k/2)}{\sqrt{t_0^2\cos^2(k/2)+\Delta^2\sin^2(k/2)}}\right).
\end{equation}
Note that the equality of the modulus of the $X$ and $Y$ matrix elements for the scattering of a Majorana zero mode to a bulk mode or a pair creation/annihilation, respectively, is not special to the model or the parameters we use here, but is deeply rooted in the topological protection of the ground state degeneracy. It is a direct consequence of the fact that zero modes based on MBSs have equal contributions of electrons and holes in their wave function.

In order to obtain a manageable expression for the fluctuation-induced coupling of Majorana modes and bulk bogoliubons [Eq. (\ref{maj_bulk_coupling})] we have assumed $\mu=0$. As shown in Appendix \ref{nonzero_mu_gw}, this assumption does not affect our final results in an essential way.

\section{Decay time}

In this section, we estimate the time scale on which the system gets excited (ground state decay time) due to the charge fluctuations  $\delta Q$ on the gate. Our final result has the form of Fermi's golden rule. However, due to the operator nature of the fluctuation $\delta Q$, care must be taken in order to treat thermal and quantum fluctuations correctly. We therefore describe the derivation in detail. 

We assume that at time $t=0$ the topological superconductor, described by Hamiltonian (\ref{tsc_hamiltonian}), is in one of its ground states $\left|\psi(t=0)\right>=\left|0\right>$. If the global parity of the TSC is even (odd), then $\langle0|b^\dagger_0b_0|0\rangle = 0$ (1). In the odd parity sector we are interested in processes $|0\rangle = b_0^\dagger |\Omega\rangle\rightarrow |k\rangle = b_k^\dagger |\Omega\rangle$, in which a zero-energy bogoliubon $b_0$ is scattered to a finite-energy bogoliubon $b_k$. $|\Omega\rangle$ is the bogoliubon vacuum. The corresponding process in the even parity sector is $|0\rangle = |\Omega\rangle \rightarrow |k\rangle =b_0^\dagger b_k^\dagger|\Omega\rangle$, where a pair of bogoliubons is created. The Hamiltonian restricted to these processes reads
\begin{equation}
H = \sum_k \epsilon_k \left|k\right>\left<k\right| + \lambda \delta Q \sum_{k\neq 0} X_{0k} \left[\left|k\right>\left<0\right| + {\rm H.c.}\right] + H_G,\label{ham_dirac}
\end{equation}
for both, the even and the odd parity sector. Dropping all terms involving matrix elements $X_{k,k'}$ with both $k,k'\neq 0$ corresponds to neglecting terms of higher order in perturbation theory in the final Fermi's golden rule result.

We now calculate the ground state decay rate of Hamiltonian (\ref{ham_dirac}) by a method based on time-dependent perturbation theory. The same result can also be derived with non-equilibrium Green's functions on the Keldysh contour, but this method is significantly more involved so that we have chosen to describe the simpler perturbation theory method here. The Ansatz for the time-dependent state of the TSC is
\begin{equation}
\left|\psi(t)\right> = \left|0\right> + \sum_{n=1}^\infty \lambda^n \sum_k a_k^{(n)}(t)\left|k\right>.
\end{equation}
However, one must keep in mind that in this formulation the coefficient functions $a_k^{(n)}(t)$ are operators in the Hilbert space of the electrons on the gate, and these operators must be averaged ($\left<\cdot\right>_G$) at some point. Thus, the usual expression for Fermi's golden rule $\D |a_k(t)|^2/\D t$ is ill defined. It is easily seen that an expression $|a_k(t)|^2$ does not respect the proper time order of the operators $\delta Q$ [see Eqs. (\ref{as_eqs_of_motion}) and (\ref{fgr1}) below]. The proper way of deriving Fermi's golden rule in the present context starts from the probability
\begin{equation}
\hat P_k(t) = \left<\psi(t) |k\right>\left<k| \psi(t)\right>
\end{equation}
that the $k$th bogoliubon is occupied. And since $\hat P_k(t)$ is itself an operator for the gate degrees of freedom, it finally must be averaged over the thermal occupation of the electrons in the gate
\begin{equation}
P_k(t) = \left<\hat P_k(t)\right>_G = \frac{\tr\left[ e^{-\beta (H_G-\epsilon_F)} \hat P_k(t)\right]}{\tr\, e^{-\beta (H_G-\epsilon_F)}},
\end{equation}
where $\beta$ is the inverse electronic temperature in the gate and $\epsilon_F$ is the Fermi level in the gate. The equations of motion for the coefficients $a_k^{(n)}(t)$ are
\begin{align}
\dot a_k^{(0)}(t) &= 0 \nonumber \\
\dot a_k^{(1)}(t) &= - \frac i\hbar \delta Q(t) X_{0k} e^{i\epsilon_k t/\hbar},\label{as_eqs_of_motion}
\end{align}
where $\delta Q(t) = e^{iH_G t} \delta Q e^{-i H_G t}$ are the Heisenberg gate-charge fluctuation operators. Using the integrals of Eqs. (\ref{as_eqs_of_motion}), the leading (second) order term in $P_k(t)$ is
\begin{multline}
P^{(2)}_k(t) = \lambda^2 \left< \left[a_k^{(1)}(t)\right]^\dagger a_k^{(1)}(t)\right>_G \\ 
= \frac{\lambda^2}{\hbar^2} |X_{0k}|^2 \int_0^t \D t_1\D t_2 \left<\delta Q(t_1)\delta Q(t_2)\right>_G e^{-i\epsilon_k(t_1-t_2)}, \label{fgr1}
\end{multline}
for $k\neq 0$. In this expression, the order of the operators $\delta Q(t_1)$ and $\delta Q(t_2)$ has been properly accounted for. The required autocorrelation function of the gate charge
\begin{equation}
C(t) = \left<\delta Q(t) \delta Q(0)\right>_G = \int \D\omega e^{-i\omega t}C(\omega)
\end{equation}
is calculated in Appendix \ref{gate_correlator}. It depends on the spatial dimensionality $D$ of the gate and on its linear size $l$. We find
\begin{align}
C_D(\omega) = B_D \exp(-\omega^2/8\epsilon_F\epsilon_{\rm cut}) \frac{\omega}{\left[1-\exp(-\beta\omega)\right]}\label{gate_spectrum}
\end{align}
with
\begin{align}
B_1 &= 1/8\epsilon_F\epsilon_{\rm cut} & B_3 &=1/8\epsilon_{\rm cut}^2.
\end{align} 
Here $\epsilon_F$ is the Fermi energy in the gate, and $\epsilon_{\rm cut} = \hbar^2/2m^* l^2\simeq 38$ meV $\frac{m_0}{m^*} ({\rm nm}/l)^2$ is an energy scale which must be much smaller than the Fermi energy for Eq. (\ref{gate_spectrum}) to be valid. $\epsilon_{\rm cut}$ plays the role of an ultraviolet cutoff below the band edge cutoff. It is introduced by the size $l$ of the gate region which is coupled to the TSC. In the following discussion, the cutoff $\exp(-\omega^2/8\epsilon_F\epsilon_{\rm cut})$ takes place at much higher energies than we are interested in, and thus it serves only as a regularization. The temperature cutoff in Eq. (\ref{gate_spectrum}), however, turns out to be much more important.

The spectral function $C_D(\omega)$ of the gate charge fluctuations is not symmetric in $\omega$. For zero temperature one finds that $C(-|\omega|)=0$, i.e., the quantum fluctuations are characterized by the positive frequency part of the spectral function. The  spectral weight for negative frequencies is non-zero only at finite temperatures, and exponentially small in $|\omega|/T$ for $\omega<0$.

From Eq. (\ref{fgr1}) one may derive Fermi's golden rule in the usual way,\cite{sakurai} and one obtains for the ground state decay rate
\begin{multline}
\hbar/\tau = 2\pi \sum_m |X_{0m}|^2 \lambda^2 C_D(-\epsilon_m) \\= 2\pi B_D \lambda^2 \sum_m |X_{0m}|^2  \exp\left(-\frac{\epsilon_m^2}{8\epsilon_F\epsilon_{\rm cut}}\right) \frac{\epsilon_m}{e^{\beta\epsilon_m}-1}. \label{dec_rate_cumb}
\end{multline}
Only the spectral weight $C_D(\omega)$ of the gate charge fluctuations with negative frequencies, i.e., the thermal charge fluctuations, enters the decay rate. Quantum fluctuations, which are present even at zero temperature, only give rise to spectral weight at positive frequencies, which is irrelevant for $\tau$.

In the continuum limit ($N\rightarrow\infty$), this rate may be calculated approximately in the limit $t_0 \gg \Delta$ and $T\gg\Delta$ (see Appendix \ref{appendix_approximate_rate})
\begin{multline}
\frac\hbar\tau \simeq \frac\pi2\lambda^2B_D T\exp(-\Delta/T) \\ \simeq \left\{\begin{matrix}\frac{10.7}{\epsilon^2\epsilon_F[{\rm eV}]}\frac{m^*}{m_0} \left(\frac{l}{d}\right)^2 T \exp(-\Delta/T) & \text{for } D=1\\
\frac{281.6}{\epsilon^2}\left(\frac{m^*}{m_0}\right)^2 \frac{l[nm]^4}{d[nm]^2} T \exp(-\Delta/T) & \text{for } D=3\end{matrix}\right. \label{main_result}
\end{multline}
This is our main result. Apparently, the rate becomes independent of $t_0$ in this limit, which is perfectly reasonable since $t_0$ is only a model-specific detail. The physically important parameters are the size of the SC gap $\Delta$ and the gate temperature $T$.

\begin{figure}[!ht]
\centering
\includegraphics[width=\linewidth]{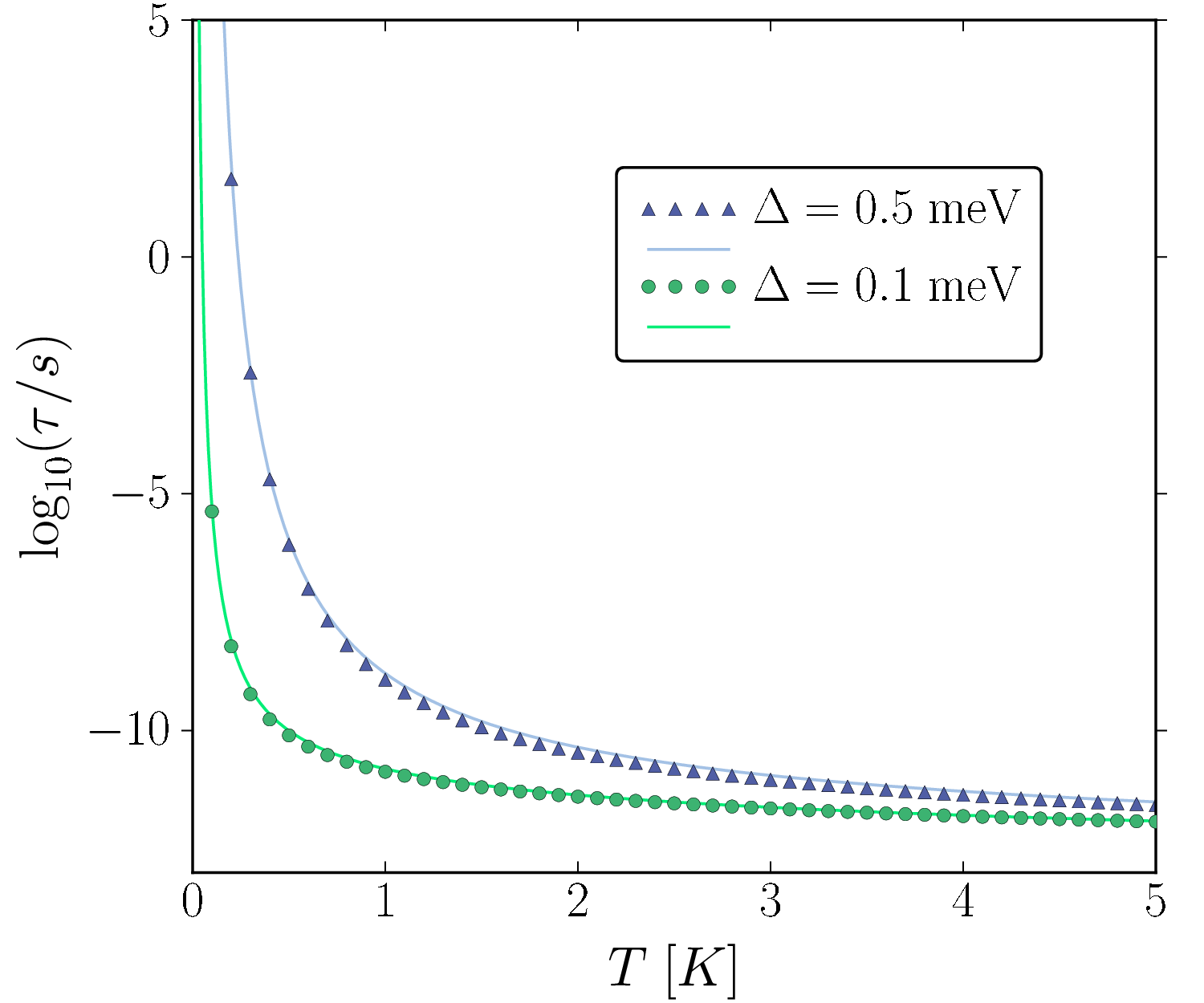}
\caption{(Color online) Decay times $\log_{10}(\tau/{\rm s})$ as a function of gate temperature in Kelvin. The upper blue curve corresponds to the case $\Delta = 0.5$ meV $\simeq 5$ K and the lower green curve to $\Delta = 0.1$ meV $\simeq 1$ K. The dots have been calculated numerically from Eq. (\ref{dec_rate_cumb}) and the lines are the approximate form (\ref{main_result}). For the numerical calculation, $t_0=10$ meV has been chosen. All other parameters have been set such that the pre-factor equals unity, $\pi\lambda^2 B_D/2=1$.}
\label{fig_rate}
\end{figure}

From Fig. \ref{fig_rate} one can see that Eq. (\ref{main_result}) is a very good approximation of Eq. (\ref{dec_rate_cumb}), even for temperatures much lower than $\Delta$. Only for $\Delta/T\simeq 0.02$ there are significant deviations from the approximate form and the pre-factor acquires a different power law in $\Delta/T$ (see Appendix \ref{appendix_approximate_rate}). However, in this regime the exponential suppression is already so strong that the power-law pre-factor is irrelevant. Thus Eq. (\ref{main_result}) can be used in all relevant parameter regimes.

\section{Discussion}

The most important parameter entering the ground state decay rate $\tau^{-1}$ in Eq. (\ref{main_result}) is the ratio of  the TSC gap $\Delta$ and the temperature $T$. For $T\ll\Delta$ the rate $\tau^{-1}$ is exponentially suppressed by a factor $\exp(-\Delta/T)$, as expected for a system in thermal equilibrium. However, the pre-factor gives hints about the optimal gate design. First of all one can see that there are huge differences between a 1D and a 3D gate. This is due to the different pre-factors $B_D$ in the correlation function of the gate charge fluctuations. For realistic parameters one finds $B_3/B_1\sim 10^3$. For a 1D gate, the pre-factor is of order one for $l=d$ (linear size $l$ of the active gate region is equal to the gate-wire distance), $\epsilon=1$, $\epsilon_F = 1$eV, and $m^* = 0.1 m_0$. For a 3D gate, the gate size $l$ does not scale with the gate-wire distance $d$. With the same parameters and $l=d=50$ nm the pre-factor is $\sim 7000$. From this one may draw the conclusion that the optimal gate has (effectively) low dimensions, a small active region, and is not too close to the TSC. Similarly, the inverse dependence on the dielectric constant $\epsilon$ and on the gate material parameters $m^*$ and $\epsilon_F$ gives further hints for the optimal gate design.

\begin{figure}[!ht]
\centering
\includegraphics[width=\linewidth]{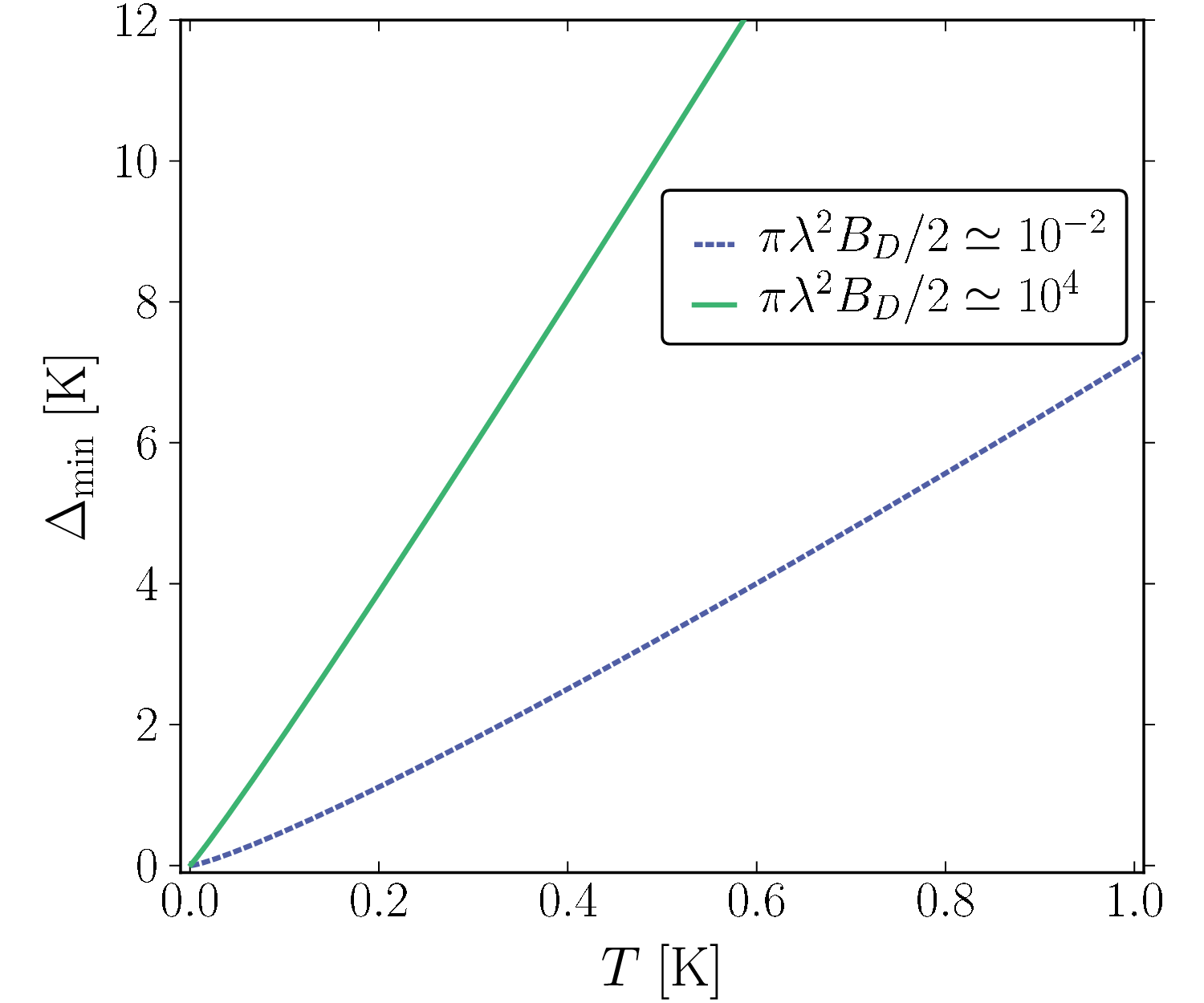}
\caption{(Color online) Minimal gap of the TSC $\Delta_{\rm min}$ in Kelvin as a function of the gate temperature $T$. The upper green (solid) line corresponds to the worst-case scenario in which the rate pre-factor $\pi\lambda^2B_D/2\simeq 10^4$, while the lower blue (dashed) line refers to the best-case scenario for which $\pi\lambda^2B_D/2\simeq 0.01$ (see the main text).}
\label{fig_limits}
\end{figure}

We are now in a position to give reasonable limits for the two important parameters $\Delta$ and $T$. For this we calculate the minimum gap $\Delta_{\rm min} \simeq T\log(T\pi\lambda^2 B_D \tau / 2\hbar)$, required for a topological protection that should last for at least one microsecond, i.e. $\tau>1\,\mu$s. Of course, $\Delta_{\rm min}$ depends on the pre-factor $\pi\lambda^2 B_D/2$, which may be tuned by several orders of magnitude. We assume here a worst case scenario $\pi\lambda^2B_D/2 =10^4$ for a 3D gate with $l=d=50$ nm, and a best case scenario $\pi\lambda^2B_D/2 =0.01$ for an optimized 1D gate (e.g., dielectric constant $\epsilon=5$ and $l/d=1/2$). Figure \ref{fig_limits} shows $\Delta_{\rm min}$ for gate temperatures smaller than 1 Kelvin and for the best and worst case scenario.

Approximately, a TSC gap $\Delta > 20~T$ is sufficient to obtain a TSC, the ground state of which is stable for at least one microsecond. By optimizations in the gate design this ratio may be reduced by a factor $\sim 4$.

\section{Conclusion}

We have studied the loss of quantum information stored in a Majorana qubit due to thermal charge fluctuations on a nearby gate, which is capacitively coupled to the topological superconductor (TSC) hosting the Majorana fermions. The fluctuating gate excites the TSC above its ground state. Since the topological protection is only active as long as the TSC is in its ground state, the stored information gets lost as soon as a finite-energy excitation is created. Starting from a tight-binding model of a TSC,\cite{kitaev} we derived a microscopic description of the gate-TSC coupling via the Coulomb interaction. Finally, we obtained an expression for the qubit lifetime which only depends on the gate temperature $T$, the superconducting gap $\Delta$, and some parameters describing the gate geometry and material parameters. Our result does practically not depend on the hopping amplitude and the chemical potential of the TSC model. 

It was found that $\Delta/T \simeq 20$ is required for a qubit lifetime of at least one microsecond, if a non-optimized gate design is chosen. By using optimized gate geometries and dielectrics, however, this ratio may be reduced by factors of 4 and more.

This work shows that not only the temperature of the topological superconductor, in which quantum information is stored, is relevant for the coherence time. Also the temperature of distant parts in the experimental setup, coupled to the superconductor via the long-ranged Coulomb interaction, is indeed important and must be sufficiently small compared to the gap of the superconductor. On top of that, we mention again that any experimental realization of a topological superconductor still suffers of the seemingly ubiquitous problem of quasiparticle poisoning.

Finally we note that thermal fluctuations are not the only relevant source of gate noise. For a functional quantum computer based on Majorana qubits it is also important to investigate the non-equilibrium noise due to the changes in the gate voltages, needed for the actual braiding operations. This issue may be studied on the basis of the model we have derived, but is beyond the scope of this work.

\acknowledgements
M.J.S. acknowledges interesting discussions with Fabian Hassler. 
D.R. and D.L. acknowledge financial support by the Swiss SNF, NCCR Nanoscience, NCCR QSIT, and the EU project SOLID.

\appendix

\section{Gate-wire coupling for $\mu\neq 0$\label{nonzero_mu_gw}}

In the main text we have restricted the discussion to $\mu=0$ in order to obtain manageable expressions for the gate-wire coupling. In this section we show that relaxing this assumption does not change the results dramatically.

The Bogoliubov transformation from the electron basis $c_n$ to the bogoliubon basis $b_k$ in the topological superconductor reads
\begin{align}
b_l &= \alpha_{l,n} c_n + \beta_{l,n} c^\dagger_n.
\end{align}
$l$ is not necessarily related to a momentum but should be thought of as a general label of the Bogoliubov eigenstates with energy $\epsilon_l$. $n$ is a real space index, labeling the sites of the TSC wire. For $\mu=0$ and $N$ odd the Bogoliubov transformation becomes particularly simple because $H_{\rm TSC}$ can be decomposed into two uncoupled chains with alternating hoppings when written formally in terms of Majorana operators
\begin{multline}
H_{\rm TSC} = i\sum_n \biggl[(\Delta + t_0)\gamma_{Bn}\gamma_{An+1} \\
+ (\Delta-t_0)\gamma_{An}\gamma_{Bn+1} - 2\mu \gamma_{An}\gamma_{Bn}\biggr],
\end{multline}
where $c_n = \gamma_{A,n}+i\gamma_{B,n}$. Moreover, because of $N$ odd, the two end Majoranas are located on different subchains and can therefore be treated separately even for finite $N$. For the zero-energy bogoliubon one finds
\begin{equation}
b_0^\dagger = \frac{\sqrt{1-\delta^2}}2 \sum_{n=0}^{(N-1)/2} \delta^n(c_{2n+1} + c_{2n-1}^\dagger + c_{N-2n}^\dagger - c_{N-2n} ),\label{b0_operator}
\end{equation}
with $\delta = (\Delta-t_0)/(\Delta+t_0)$ and for the bulk bogoliubons, where the index is $l=(k,s)$ with $s=1,2$ for the two subchains, we have
\begin{align}
b_{k,1}^\dagger &= \sum_{n=1}^{(N+1)/2} \biggl[ \sin(k n)(c_{2n} - c_{2n}^\dagger)\nonumber \\
& \bs\bs- \sin(kn + \alpha^1_k)(c_{2n-1} + c_{2n-1}^\dagger)\biggr] \\
b_{k,2}^\dagger &= \sum_{n=1}^{(N+1)/2} \biggl[ \sin(k n)(c_{2n} + c_{2n}^\dagger) \nonumber \\
& \bs\bs + \sin(kn + \alpha^2_k)(c_{2n-1} - c_{2n-1}^\dagger)\biggr].\label{bk2_operator}
\end{align}
The corresponding energies are given in Eq. (\ref{bog_energy}) and are equal for both subchains.

Knowing the Bogoliubov transformation matrices, which can be read off Eqs. (\ref{b0_operator})-(\ref{bk2_operator}), one may rewrite the gate-wire coupling $H_c$ in terms of $b_l$ [see Eq. (\ref{gw_coupling_bogol})]. Straightforward algebra shows that the matrix elements $X_{0l}$ of $H_c$ are given by
\begin{align}
X_{l,l'} &= \alpha_{l,n} F_{n} \alpha_{l',n}^* - \beta_{l,n} F_{n} \beta_{l',n}^* \label{Xme}  \\
Y_{l,l'} &= \beta^*_{l,n} F_{n} \alpha_{l',n}^* - \alpha_{l,n}^* F_{n} \beta_{l',n}^* .
\end{align}
If one of the indices $l$ or $l'$ corresponds to the zero mode, and this is the only case we are interested in, then one can easily show that $|X_{0,l}| = |Y_{0,l}| = |X_{l,0}| = |Y_{l,0}|$ and it is thus sufficient to consider $X_{0,l}$. Furthermore, since $\alpha_{0,n}$ and $\beta_{0,n}$ are exponentially small for $n$ away from one of the ends, we set $F_n=1$ for $n< N/2$ and zero otherwise. This corresponds to a gate which affects the TSC over a region which is at least as large as the spatial size of the MBS wave function. The latter is typically on the order of 100 nm. Since the wave function of the MBS ($\alpha_{0,n}$ and $\beta_{0,n}$) is exponentially small for $n\simeq N/2$ where $F_n$ deviates from unity, the particular form of the profile function $F_n$ near $N/2$ is not important and $F_n$ can be chosen to be technically most convenient. Evaluating Eq. (\ref{Xme}) under those assumptions (i.e., $\mu=0$; $N$ odd; $F_n=1$ for $n<N/2$ and 0 otherwise) results in the simple analytical expression Eq. (\ref{maj_bulk_coupling}). Note that the $X_{0,l}$ is only non-zero if $l=(k,2)$, i.e., for the second subchain.

\begin{figure}[!ht]
\centering
\includegraphics[width=\linewidth]{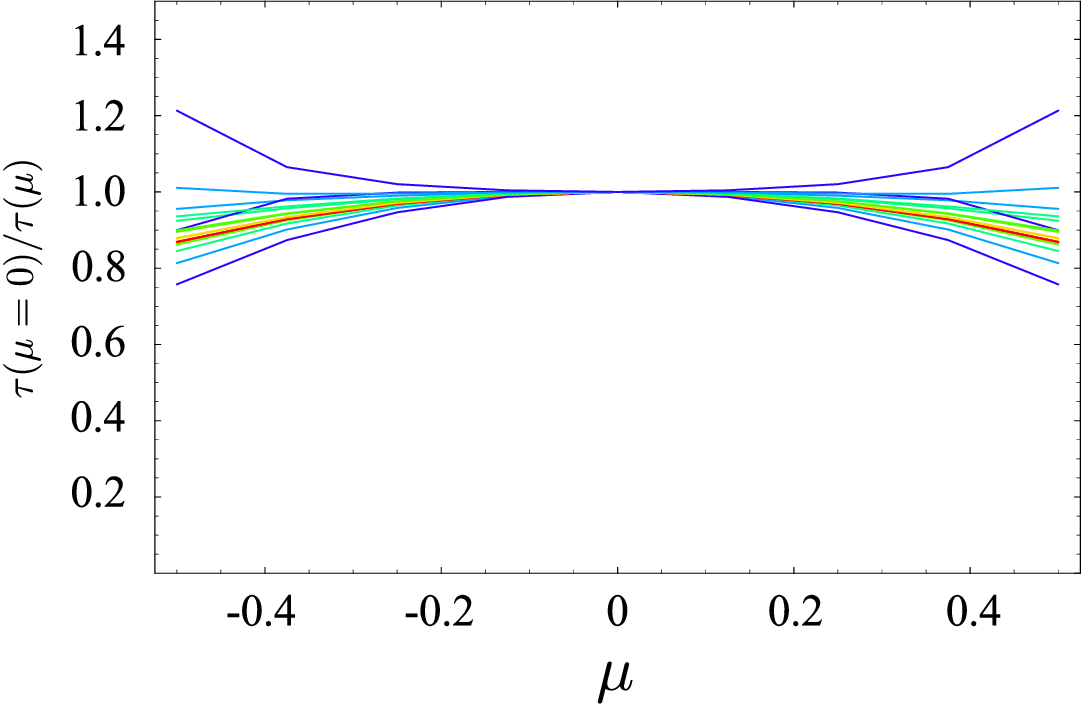}
\caption{(Color online) The ratio $\tau(\mu=0) / \tau(\mu) $ of the scattering rates at different chemical potentials $\mu$. We used the parameters $\Delta = 0.1, 0.2, ...,0.6$, $t_0=1$, $T=\Delta^*(\mu), \Delta^*(\mu)/10, \Delta^*(\mu)/100$, $\epsilon_F=4$ and $\epsilon_{\rm cut} = 0.004$. The number of sites in the TSC wire in this calculation is $N=501$. The result of a calculation with $N=500$ cannot be distinguished from this figure. The curves with larger deviations from 1 correspond to larger $\Delta$ and smaller temperatures.}
\label{fig_scat_rate_comp}
\end{figure}

In the remainder of this section we calculate the Bogoliubov transformation $\alpha_{l,n}$ and $\beta_{l,n}$ numerically from the eigenvectors of the TSC wire Hamiltonian $H_{\rm TSC}$ in Eq. (\ref{kitaev_hamiltonian}) with an arbitrary number $N$ of sites and for general $\mu$. For $\mu\neq 0$ the subchains are coupled. From $\alpha_{l,n}$ and $\beta_{l,n}$ we calculate $X_{0,l}$ numerically and use it to evaluate the Fermi's golden rule expression for the scattering rate $1/\tau$ [Eq. (\ref{dec_rate_cumb})]. The main effect of a nonzero chemical potential on the scattering rate is a trivial renormalization of the superconducting gap
\begin{equation}
\Delta \rightarrow \Delta^*(\mu) = \Delta \sqrt{1-\frac{\mu^2}{t_0^2-\Delta^2}}.
\end{equation}
Thus, in order to resolve the nontrivial correction, we calculate the scattering rate for $\mu$-dependent temperatures $T=\Delta^*(\mu)/M$, with $M=1,10,100$, and compare the result to the scattering rate at $\mu=0$. Figure \ref{fig_scat_rate_comp} shows that $\tau$ depends only weakly on the chemical potential. Typically, the scattering rate is renormalized by factors between $\frac12$ and $\frac32$ for $|\mu|>0$. Note also that the chemical potential is varied over a relatively large range $\sim t_0$ in Fig. \ref{fig_scat_rate_comp}.

Furthermore, we have calculated Fig. \ref{fig_scat_rate_comp} for $N=500$ and $N=501$. The difference is not visible. Thus, the simplifying assumptions $N$ odd and $\mu=0$ are well justified.

\section{Gate correlation functions\label{gate_correlator}}
In this appendix we outline the calculation of the correlation function of the gate charge fluctuations $C(t) = \langle \delta Q(t) \delta Q(0)\rangle_G$, where
\begin{equation}
\langle\cdots\rangle_G = \frac{\tr \left[ e^{-\beta (H_{\rm G}-\epsilon_F)} \cdots\right]}{\tr\, e^{-\beta (H_{\rm G}-\epsilon_F)}}.
\end{equation}
In frequency space and $D$ dimensions we have
\begin{multline}
C_D(\omega) = \left(\frac{l^2}{4\pi}\right)^D \int \D^D\ve k \D^D\ve q e^{-E_{\ve q}/2\epsilon_{\rm cut}} \\\delta(\omega-E_{\ve k}+E_{\ve k+\ve q}) n_{\ve k+\ve q} (1-n_{\ve k}),\label{general_correlator_integral}
\end{multline}
with the Fermi factors
\begin{equation}
n_{\ve k} = (\exp(\beta(E_{\ve k}-\epsilon_F))+1)^{-1},
\end{equation}
and the electronic energy in the gates
\begin{equation}
E_{\ve k} = \frac{\ve k^2}{2m^*}.
\end{equation}
Furthermore, the cutoff energy scale
\begin{equation}
\epsilon_{\rm cut} = \frac{1}{2m^* l^2}
\end{equation}
has been introduced. We now calculate the integral (\ref{general_correlator_integral}) for one and three spatial dimensions.
\subsection{$D=1$}
We first eliminate the $k$ integral via the Dirac delta function and transform the $q$ integration to an energy integration $E=E_q$. This gives
\begin{equation}
C_1(\omega) = \frac1{8\pi\epsilon_{\rm cut}} \int_0^\infty \frac{\D E}{E} e^{-E/2\epsilon_{\rm cut}} \mathcal P\left[\omega,\frac{\omega^2}{4 E} + \frac E 4 - \epsilon_F,T\right],\label{b6}
\end{equation}
with the smeared plateau function 
\begin{multline}
\mathcal P(w,x,T) = (e^{-(x+w/2)/T}+1)^{-1} (e^{(x-w/2)/T}+1)^{-1}
\end{multline}
describing a plateau with height 1 and width $w$ centered around $x=0$. The plateau edges are smeared over $\Delta x\sim T$.

The second argument of the plateau function in Eq. (\ref{b6}) is close to zero for $E_1=4\epsilon_F$ and for $E_2=\omega^2/4\epsilon_F$, where we have assumed that $\omega\ll\epsilon_F$. Because of the exponential cutoff $\epsilon_{\rm cut}\ll\epsilon_F$ we may neglect the first energy region. We do this by pulling the exponential function out of the integral at $E=E_2$ and at the same time regularizing the integral such that the integrand is exponentially small for energies near $E_1$. The simplest way to do this is to change the sign of the $E/4$ term in Eq. (\ref{b6}). This does not affect the integral for energies around $E_2$ but changes the sign of $E_1$. Thus, we have
\begin{widetext}
\begin{align}
C_1(\omega) &\simeq \frac{1}{8\pi\epsilon_{\rm cut}} \exp\left(-\frac{\omega^2}{8\epsilon_F\epsilon_{\rm cut}}\right) \int_0^\infty \frac{\D E}{E}  \mathcal P\left[\omega,\frac{\omega^2}{4 E} - \frac E 4 - \epsilon_F,T\right] \\
&= \frac{1}{8\pi\epsilon_{\rm cut}} \exp\left(-\frac{\omega^2}{8\epsilon_F\epsilon_{\rm cut}}\right) \int_{-\infty}^\infty\frac{2\D y}{E+\omega^2/E} \frac{e^{\omega/2T}}{\cosh(\omega/2T)+\cosh((y-\epsilon_F)/T)}\\
&\simeq \frac{1}{8\pi\epsilon_{\rm cut}} \exp\left(-\frac{\omega^2}{8\epsilon_F\epsilon_{\rm cut}}\right)\frac{2}{4\epsilon_F}\int_{-\infty}^\infty \D y \frac{e^{\omega/2T}}{\cosh(\omega/2T)+\cosh((y-\epsilon_F)/T)} \\&= \frac{1}{8\pi\epsilon_F\epsilon_{\rm cut}} \exp\left(-\frac{\omega^2}{8\epsilon_F\epsilon_{\rm cut}}\right) \frac{\omega}{1-\exp(-\omega/T)},
\end{align}
\end{widetext}
where we have introduced $y=\omega^2/4E-E/4$. Furthermore we have used that the integrand is only large for $E\simeq E_2$. All approximations become exact in the limit $\epsilon_F\rightarrow\infty$. For typical $\epsilon_F\sim$ eV the relative errors of the approximations used are on the order of $10^{-5}$.
\subsection{$D=3$}
For a gate with three spatial dimensions we first eliminate the integral over the angle between the vectors $\ve k$ and $\ve q$ via the Dirac delta function and obtain
\begin{multline}
C_3(\omega) =  \frac{l^6 m^*}{8\pi} \int_{k_-}^\infty \D k\, k\, n(E_k-\omega)(1-n(E_k))\\ \int_{q_-}^{q_+} \D q\,q\,e^{-l^2q^2/2},
\end{multline}
with $k_-={\rm Re}\,\sqrt{2m\omega}$ and $q_\pm = |k\pm\sqrt{k^2+2m\omega}|$. With $y=k^2/2m^*-\omega/2$ one may write
\begin{multline}
C_3(\omega) = \frac1{16\pi\epsilon_{\rm cut}^2} \\\int_{|\omega/2|}^\infty \D y \frac{\exp(-y/\epsilon_{\rm cut})\sinh(\sqrt{y^2-\omega^2/4}/\epsilon_{\rm cut})}{\cosh(\omega/2T)+\cosh((y-\epsilon_F)/T)}.
\end{multline}
For $\epsilon_F\gg T,\omega,\epsilon_{\rm cut}$ this integral may be approximate as described above in the $D=1$ case and one obtains
\begin{equation}
C_3(\omega) = \frac1{8\pi\epsilon_{\rm cut}^2}\exp\left(-\frac{\omega^2}{8\epsilon_F\epsilon_{\rm cut}}\right) \frac{\omega}{1-\exp(-\omega/T)}.
\end{equation}

\section{Approximative evaluation of the excitation rate\label{appendix_approximate_rate}}

In this appendix we outline the approximation of the Fermi's golden rule expression of the excitation rate given in Eq. (\ref{dec_rate_cumb}). In the limit $N\rightarrow\infty$ the sum over $m$ becomes a $k$ integral and $\tau^{-1}$ is proportional to
\begin{equation}
I = (1-\delta^2)\int_0^\pi \D k \frac{(\sin(k+\alpha_k)-\delta \sin\alpha_k)^2}{(1+\delta^2-2\delta\cos k)^2}\frac{\epsilon_k}{e^{\epsilon_k/T}-1}.
\end{equation}
After transforming the integral to $z=t_0(\pi - k)/\Delta$, the limit $t_0/\Delta\rightarrow\infty$ may be performed for $I$. We define
\begin{multline}
I^* = \lim_{t_0\rightarrow\infty} I \\
=32\Delta \int_0^\infty \frac{z^2 \D z}{(4+z^2)^{5/2}(\exp(\frac\Delta{2T}\sqrt{4+z^2})-1)}.\label{c2}
\end{multline}
In realistic situations, $t_0$ may be assumed to be at least two orders of magnitude larger than the induced superconducting gap $\Delta$. In this case, the relative error $(I-I^*)/I$ is on the order of $10^{-4}$. Thus, using $I^*$ is an excellent approximation. In this way, the hopping parameter $t_0$ of the TSC model, which has not much direct physical meaning, is removed from the rate. 

\begin{figure}
\centering
\includegraphics[width=\linewidth]{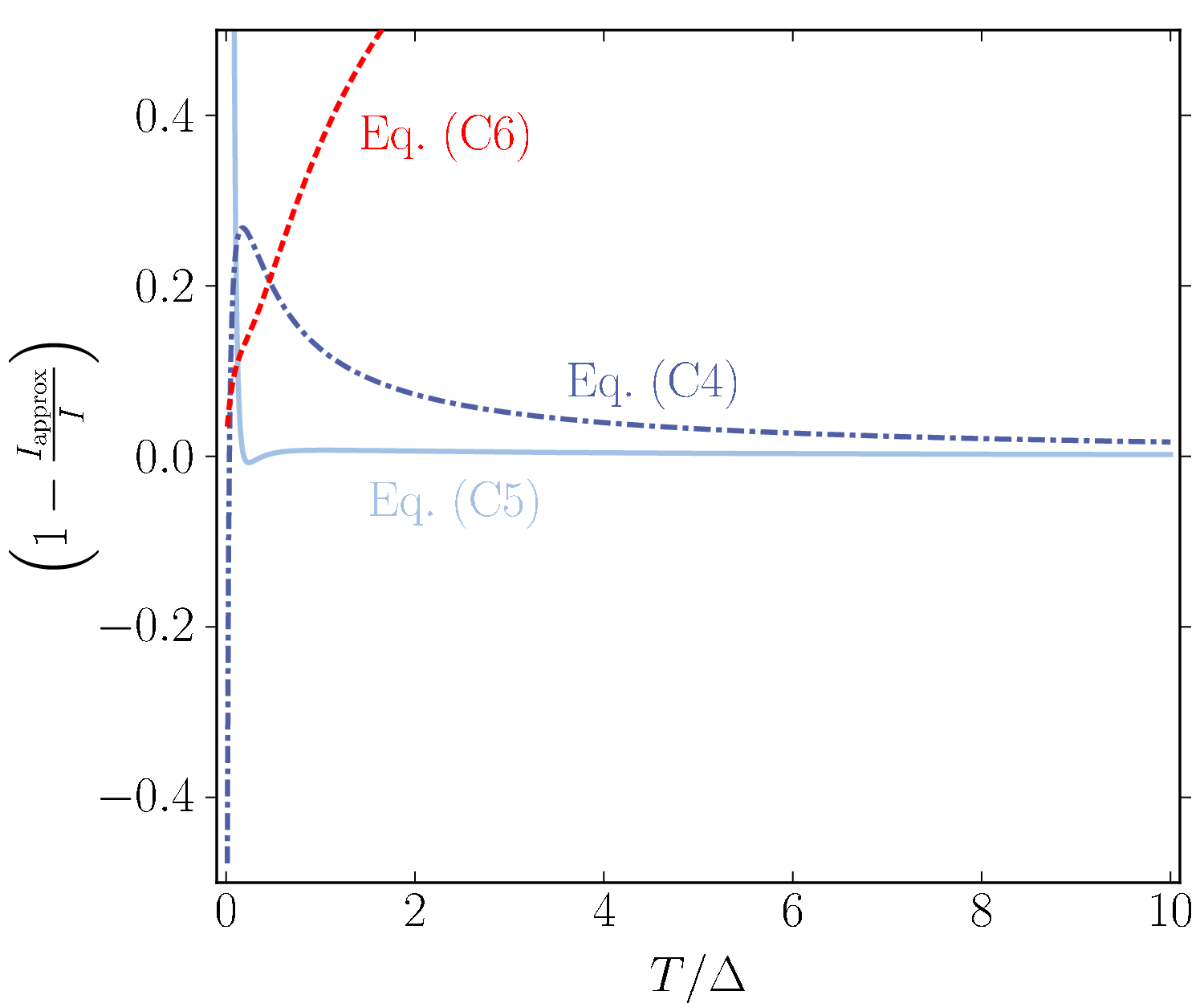}
\includegraphics[width=\linewidth]{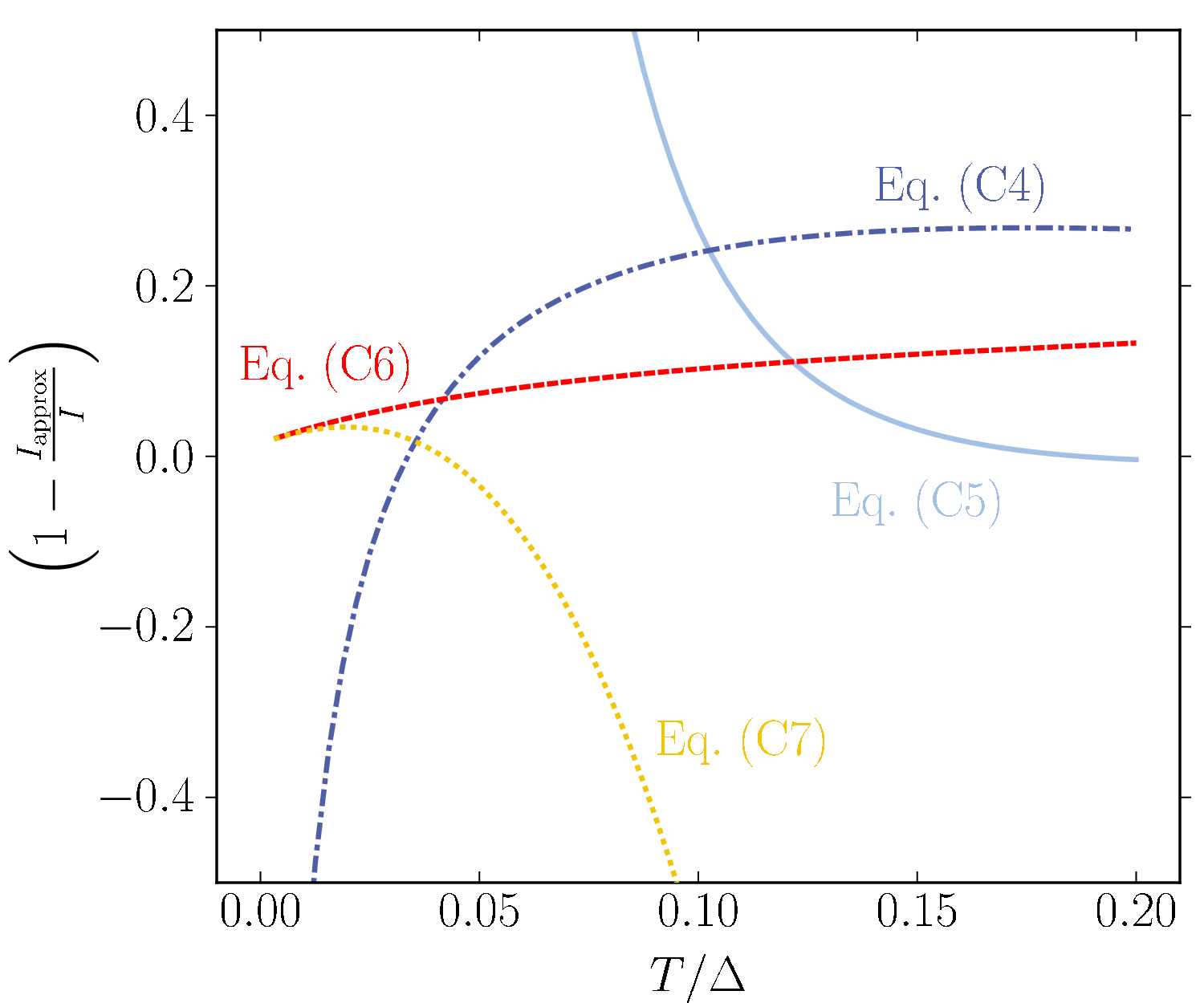}
\caption{(Color online) Relative errors $(1-I_{\rm approx}/I)$ for various approximate expressions $I_{\rm approx}$ of the rate integral $I$ as a function of $T/\Delta$. Part (b) is a zoom of part (a) for small values of $T/\Delta$.}
\label{fig_approx}
\end{figure}

The expression (\ref{c2}) is dominated by the exponential factor $\exp(-\Delta/T)$. We thus continue by evaluating the $\Delta/T\rightarrow0$ limit of $I^* e^{\Delta/T}$
\begin{equation}
\lim_{\Delta/T\rightarrow0} I^*e^{\Delta/T} = 64 T\int_0^\infty \D z \frac{z^2}{(4+z^2)^3} = \frac{\pi T}2,
\end{equation}
so that for $\Delta\ll T$
\begin{equation}
I^* \simeq \frac{\pi T}2\exp(-\Delta/T). \label{c4}
\end{equation}
Note that $\Delta/T=0$ is a singular point of $I^*$ so that the expansion in $\Delta/T$ can be performed up to the fourth order. However, this expansion has zero convergence radius. It becomes better for $\Delta/T\lesssim10$ up to second order and then becomes worse again. In second order one finds
\begin{equation}
I^* \simeq \left(\frac\pi 2+\left(\frac\pi2-\frac 43\right)\frac\Delta T+\left(\frac{5\pi}{12} -\frac 43\right)\frac{\Delta^2}{T^2} \right) T e^{-\Delta/T}. \label{c5}
\end{equation}
For $\Delta/T>10$ this approximation is much worse than the simple limit (\ref{c4}).

Next, we evaluate $I^*$ in the limit of large $\Delta/T$. In this case the exponential function is much larger than unity so that the 1 may be neglected in Eq. (\ref{c2}). Furthermore one may expand the square root in the argument. One finds
\begin{multline}
I^*\simeq 32\Delta \int_0^\infty \frac{z^2 \D z}{(4+z^2)^{5/2}} \exp(-\frac\Delta{T}(1+z^2/8)) \\= 2\sqrt\pi U(3/2;0;\frac{\Delta}{2T})\Delta e^{-\Delta/T}, \label{c6}
\end{multline}
with Tricomi's confluent hypergeometric function $U(a;b;z)$. From Fig. \ref{fig_approx} one can see that this approximation becomes worse than (\ref{c4}) for $T/\Delta \gtrsim 1/2$. The dashed line in part (b) of Fig. \ref{fig_approx} shows an expansion of $U(3/2;b;z)$ up to order $(T/\Delta)^{7/2}$, i.e.
\begin{equation}
\sqrt{\frac{\pi}{2}} \left[8 \left(\frac T\Delta\right)^{\frac32} -60 \left(\frac T\Delta\right)^{\frac52} +525 \left(\frac T\Delta\right)^{\frac72}\right] \Delta e^{-\Delta/T}.
\end{equation}
This expansion is only valid for extremely low temperatures where the exponential suppression dominates the rate.

Thus we may conclude that, although (\ref{c6}) is a very good approximation for $T/\Delta\lesssim 0.12$ and (\ref{c5}) for $T/\Delta \gtrsim0.12$, the simplest form (\ref{c4}) is sufficiently accurate in the relevant range down to $T/\Delta\simeq 0.02 $. Below this temperature the exponential suppression of the rate is so strong that the pre-factor is irrelevant.

\bibliography{mscpt}

\end{document}